\documentclass[aps,twocolumn,groupedaddress]{revtex4}

\usepackage{graphicx}

\begin{document}
\bibliographystyle{apsrev}


\title{Bose-Einstein condensation of metastable helium: some 
experimental aspects}



\author{C.I. Westbrook, A. Robert, O. Sirjean, A. Browaeys, D. 
Boiron, A. Aspect}
\homepage[]{http://atomoptic.iota.u-psud.fr/}
\affiliation{Laboratoire Charles Fabry UMR 8501 du CNRS, B.P. 147, 
91403 Orsay CEDEX, France}


\date{\today}

\begin{abstract}
We describe our recent realization of BEC using metastable helium. 
All detection is done with a microchannel plate which detects the 
metastables or ions coming from the trapped atom cloud. 
This discussion emphasizes some of the diagnostic 
experiments which were necessary to quantitatively analyze our results.
    
\end{abstract}


\maketitle

%

\section{Introduction}

Since the announcement of the observation of Bose-Einstein 
condensation (BEC) in a rubidium vapor at the 12th ICOLS meeting in 
1995\cite{cornell:95}, this fascinating state of matter has occupied central 
stage in the field of atomic, molecular and optical physics.
The many new advances reported at this meeting indicate that the field 
may continue to do so for some time to come. This paper concerns one 
of these advances, the condensation of metastable 
helium atoms (He*), and is intended as a supplement to the recently 
published Ref.~\cite{Robert:01}. 
We shall not repeat the data from that paper, but rather 
concentrate on some details that were left out of Ref.~\cite{Robert:01} 
for lack of space.

Until recently BEC had been observed in 
4 different atomic
species, H, Li, Na and Rb\cite{Anderson,Davis,Bradley,Fried},
and the first question to ask before embarking 
on the quest for BEC of He* was whether a new atom was of interest. 
During the 1990s, several groups have been working on laser cooling 
of He*, and of course one answer to the above question is simply that 
the attainment of BEC is the best cooling one can do, and 
many of the same justifications for laser cooling apply to BEC. 
In addition, one might hope that a new atomic species might allow one 
to observe new phenomena, not accessible to the previously studied 
cases. 
In this respect it seemed clear that the metastability of the atoms 
might be very important. 
Simple, rapid and efficient detection of the He* atoms is possible using
electron multipliers such as
microchannel plates (MCP) and these detectors can also be used to 
observe ions resulting from Penning ionizing collisions, either with 
residual gas atoms or between the He* atoms themselves. 
Thus, when BEC was observed in alkali gases, groups working on laser 
cooling of He* naturally considered the feasibility of 
He*\cite{Vassen}.

A potential impediment to the achievement of the high 
densities necessary for evaporative cooling and BEC is the Penning 
ionization reaction:
$$
He^{*} + He^{*} \rightarrow He + He^{+} + e^{-}.
$$
Many experiments have shown that in a magneto-optical trap (MOT), this 
process is very rapid and limits the density in such a 
trap\cite{Bardou:92,Mastwijk:98a,Tol:99a,Kumakura:99a,Browaeys:99,Pereira:01}. 
Thus, when loading atoms from a MOT into a non-dissipative trap, there was a 
danger that the density and more importantly the elastic collision 
rate would be too small to efficiently cool by evaporation. 
In addition, it was known 
experimentally\cite{Mastwijk:98a,Tol:99a,Kumakura:99a} and 
theoretically\cite{Venturi:00} that even in the absence of 
resonant light, the rate constant for Penning ionization was on the order of 
$10^{-10}\:\mathrm{cm}^{3}/\mathrm{s}$. 
Thus, even if it were possible to begin evaporative cooling, such a 
large inelastic collision rate is likely to prevent evaporative 
cooling down to BEC.

On the other hand it was predicted\cite{Fedichev,Venturi} that the elastic
scattering length $a$ would be quite large for He*. 
This result was very encouraging because it
indicated that despite the low densities achievable in a MOT,
thermalizing collisions in a MOT-loaded magnetic trap could be
rapid enough to allow evaporative cooling.

Even more importantly, it was 
also known that, for a spin polarized
sample, the Penning ionization rate is suppressed by spin 
conservation.
He* has total angular momentum one, therefore only
one trapping state exists and
magnetically trapped He* atoms are neccesarily 100\%
polarized.
So one might hope to accumulate large densities
of He* in a magnetic trap. 
Experimentally, a suppression of one order of magnitude had already 
been demonstrated as early as 1972\cite{Hill:72}, 
this result was followed up by more recent 
measurements\cite{Herschbach:00,Nowak:00}.
In the mid 1990's it was predicted that the degree of 
suppression could be as high as 5 orders of 
magnitude\cite{Fedichev,Venturi}, which would easily permit long
storage times at densities necessary for BEC. 
The theoretical predictions for the scattering properties of
He* motivated the serious 
attempt to achieve BEC using a strategy analogous to that 
used in the alkali gases.

\section {Experiment}

In our apparatus, a Zeeman cooled atomic beam loads a MOT which, after 
an optical molasses cooling and an optical pumping stage,
loads a magnetic trap. 
The magnetic trap is of the cloverleaf design\cite{Mewes:96}. 
The only unusual feature of the apparatus is an MCP
placed 5 cm below the trap center. 
A grid in front of the MCP allows one to either attract or repel 
positive ions. The front face of the MCP is at negative high voltage, 
and so electrons are never detected.
We typically trapped $3\times 10^{8}$ atoms in the MOT, and 
transferred approximately 50\% to the magnetic trap.
The MOT temperature was typically of order 1 mK, while after molasses 
cooling, the atomic temperature reached 300 $\mu\mathrm{K}$.
After loading the magnetic trap, 
the atomic sample was compressed by 
lowering the magnetic field bias. 
During this process the temperature increased again to about 1 mK.
Then, an RF-knife was applied and ramped down from 130 MHz 
to effect the forced evaporation. 

\begin{figure}
\includegraphics{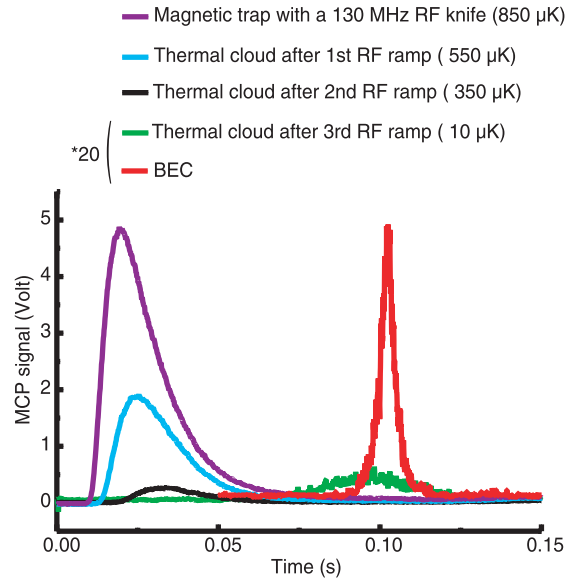} 
\caption{MCP time of flight signals at different stages of the RF 
evaporation. The vertical scales on last two curves were 
multiplied by 20. At temperatures above 100 $\mu\mathrm{K}$,
the atomic cloud expands nearly isotropically, and the MCP detects 
primarily the downward going atoms. 
At temperatures low compared to 100$\mu\mathrm{K}$, 
the atoms fall in the earth's gravititional field and are nearly all 
detected. 
The mean arrival time is about 100 ms, the time to fall 5 cm.
Because of the loss of atoms during evaporative cooling,
and the small solid angle of the detector,
very few atoms are detected at temperatures between
300 and 30 $\mu\mathrm{K}$. \label{fig:tdvrf}}
\end{figure}

The evolution of the temperature, phase space density and elastic 
collision rate during the evaporation ramp were all derived from
measurements of the time of flight distribution 
of atoms to the MCP after the magnetic trap was turned off.
Several examples are shown in Fig. 1. 
The change in gravitational energy of the atoms in falling 5 cm to 
the detector corresponds to 240 $\mu\mathrm{K}$.
For the initial temperature of the trapped atoms, this energy is
negligible compared to the kinetic energy, and so the atoms expand 
nearly isotropically after release, and the collection efficiency
of the detector corresponds roughly to its solid angle of 0.5\%.
In this situation, the time of flight distribution is peaked at a 
value corresponding to the flight time of an atom moving toward the
detector with the most probable speed at that
temperature. 
The signature for cooling is a shift of the arrival time toward later 
times as can be seen in Fig. \ref{fig:tdvrf}. 
Unfortunately, as the atoms are evaporatively cooled, their number 
also diminishes and the detected signal drastically decreases. 
Indeed when the temperature is of order 100 $\mu\mathrm{K}$, no signal 
is visible on the detector.
As the temperature decreases further however, the atoms begin to fall 
down
rather than fly away, and the collection efficiency of the detector 
increases dramatically.
Near a temperature of 10 $\mu\mathrm{K}$ we observe a ``revival'' of 
the MCP signal, and at 1 $\mu\mathrm{K}$, close to the BEC threshold, 
nearly all the atoms remaining in the trap
reach the detector and we observe the 
characteristic structure of an expanding cloud below the BEC 
temperature: a broad peak whose
width corresponds to the temperature and a narrow peak on 
top of it. 

As is discussed in Ref.~\cite{Robert:01}, a careful analysis of the narrow
BEC peak reveals that its width increases as the 1/5 power of the 
number of atoms $N_{0}$ in the peak as predicted by the Gross-Pitaevski 
equation in the Thomas-Fermi approximation. 
Upon closer examination however, the results were puzzling. 
First, it was surprising that we detected the atoms at 1 $\mu\mathrm{K}$ 
at all. A magnetic field gradient as small as 30 mG/cm is enough to 
deviate the atomic trajectory so as to miss the MCP. 
We were virtually certain that residual field gradients larger than 
this were present in the apparatus.
Secondly, the study of the expansion of the gas depends on being able 
to turn the trap off suddenly compared to the inverse of the oscillation frequencies 
in the trap (50 and 1300 Hz). 
Here again we were certain from magnetic field measurements
that effects such as eddy currents in the 
reentrant flanges holding the magnetic trap coils, limited our field 
turn-off time to about 1 ms.
Thirdly, we could estimate the number of atoms in the 
cloud at or near the critical temperature and compare it with the 
expected number in the ideal gas limit. 
The predicted number is given by\cite{Dafolvo}:
\begin{equation}\label{eq:tc}
    N_{\textrm{c}}=1.202\,(kT_{\textrm{c}}/\hbar\tilde{\omega})^{3}.
 \end{equation}
 
\noindent Here $\tilde{\omega}$ denotes the geometric mean of the trap
oscillation frequencies.
Our calibration of the detector indicated a number of atoms smaller than 
this by an order of magnitude.
Finally a quantitative analysis of the $N_{0}^{1/5}$ dependence, resulted 
in an estimate of the scattering length $a$ of order 100 nm.
Such a large value of the scattering length seemed to be ruled out by 
the fact that the condensate had a lifetime of a few s.
One expects the 3 body loss rate to scale as the fourth power of 
$a$\cite{Fedichev:96}, 
and therefore 3 body losses would have caused the BEC to decay in 
much less than 1s. 

\begin{figure}
\includegraphics{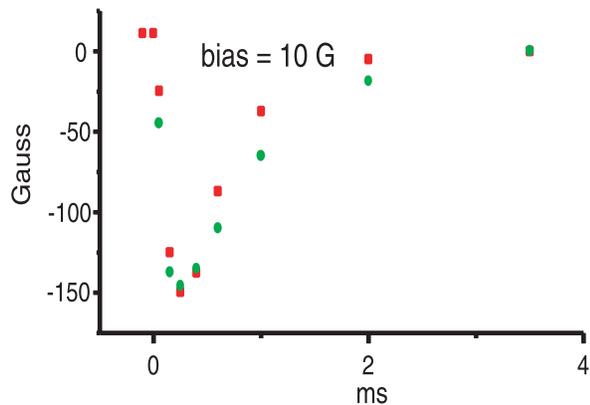} 
\caption{\textit{In situ} measurement of the magnetic field as a function of 
the time after turning off the magnetic trap. 
Black squares correspond to circular, grey circles to 
linear laser polarization.
The field rapidly reverses due to differential eddy currents. 
This reversal causes approximately 10\% of the atoms to transfer to 
the field-insensitive $m=0$ state.  \label{fig:b-field}}
\end{figure}

A clue to resolving these difficulties came from earlier, \textit{in 
situ} measurements 
of the magnetic fields in our apparatus when the trap was turned off. 
In Fig.~\ref{fig:b-field} we show the results of such a measurement. 
We began with atoms in a magnetic trap at 1 mK.
While monitoring the MCP signal, we directed a 10 mW/cm$^{2}$  
laser pulse of 20 $\mu\mathrm{s}$ duration at the cloud at a time 
$t$ after the magnetic field turnoff. 
When the laser, which propagated parallel to the bias field,
was resonant with the atoms, including the Zeeman shift 
due to the trapping fields, the atoms scattered the laser light and 
were pushed from the path to the detector. 
The laser detuning which minimized the MCP signal therefore 
corresponded to the Zeeman shift at the time of the pulse.

The magnetic field deduced from this Zeeman shift is plotted 
in Fig. \ref{fig:b-field} as a function of $t$. 
Negative times correspond to atoms still in the trap, and the value we 
observe agrees with the calculated magnetic field bias in the trap. 
One sees that the magnetic field in the trapping region undergoes a 
violent reversal during the turnoff, reaching a value of more than 
$-150\:\mathrm{G}$ in less than 100 $\mu\mathrm{s}$.
This reversal is possibly due to the combined effects of the eddy currents 
induced by the pinch coils and the compensation coils.
These two sets of coils carry currents in opposite directions and have
very different sizes. The eddy currents they induce therefore have 
different signs and probably 
different time constants, which would explain the observed behavior. 
An uncompensated trap does not exhibit a magnetic field reversal. 

It appears therefore that when the field reverses sign,
some of the atoms 
undergo transitions to the field insensitive $m=0$ state.
We presume that after the transition the atoms adiabatically 
remain in the $m=0$ state in the presence of the weak residual fields 
remaining in the apparatus. 
They therefore remain insensitive to any further field gradients.

\begin{figure}
\includegraphics{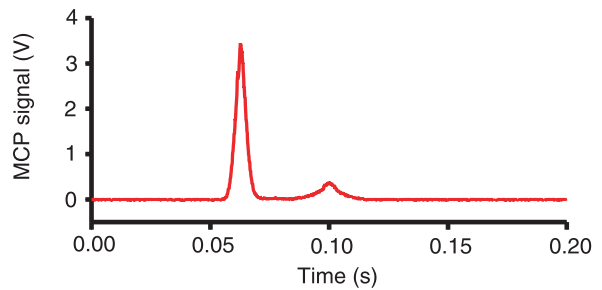} 
\caption{Time of flight spectrum in the presence of a magnetic field 
gradient of an evaporatively cooled cloud of 
atoms. The height and arrival time of the small peak are independent of 
the applied gradient. The large peak's arrival time decreases as the 
vertical gradient (about 10 G/cm) is increased. A 1 G/cm horizontal 
gradient was also applied
in order to maximize the number of atoms in the large peak. 
The ratio of the peak areas is 7.
Thus we 
believe that the small peak corresponds to atoms in the $m=0$ state,
while the 
large peak corresponds to atoms in the $m=1$ state.\label{fig:m1m0}}
\end{figure}

We can test this interpretation by deliberately applying a magnetic 
field gradient with an external coil, and observing the time of flight 
spectrum. 
The results are shown in Fig. \ref{fig:m1m0}. 
The gradient was turned on about $100\:\mu\mathrm{s}$ after the
currents to the magnetic trap were turned off. 
The figure shows two peaks, one arriving at approximately 100 ms after 
the field turnoff, as occurs without an applied gradient. 
The location and height of this peak does not depend on the magnitude 
of the gradient. 
The other peak occurs earlier, corresponding to atoms accelerated by 
the gradient and the arrival time shifts with the magnitude of the 
gradient.
The height of the peak can be varied by varying the horizontal components 
of the gradient.  
The figure shows the largest early peak we were able to produce. 
These data show that indeed a fraction of the atoms makes the flight 
to the detector in the $m=0$ state and that at least 7 times more 
atoms are in a field sensitive state after the trap turnoff.
The applied gradient was produced by a coil above the trap and thus
the atoms which are accelerated are weak field seekers ($m=+1$). 
It is also possible that some atoms are in the strong field 
seeking state ($m=-1$), but since they are accelerated upwards, they have 
little chance to reach the MCP. 

To get an estimate of the number of atoms in the magnetic trap
we can use an analysis similar to that which leads to Eq. \ref{eq:tc} leading to the number
$N_{\textrm{th}}$
of atoms in the thermal cloud below the critical temperature.
By fitting the wings of the time of flight spectra, we are able to 
determine the
temperature $T$ of
the atomic cloud, and to infer $N_{\textrm{th}}$. As discussed in 
Ref.~\cite{Dafolvo}, and experimentally
demonstrated in Refs.~\cite{Mewes:96,Hau,Ensher}, this number should be given by:
$N_{\textrm{th}}=1.202\,(kT/\hbar\tilde{\omega})^{3}$.
This relation gives an absolute thermodynamic
measurement of the number of atoms.
It is higher by a factor $f=8\pm 4$ than the value
we derive from the calibration of the MCP.
Taking this correction into account, the largest condensate
we have observed contained about $10^{5}$ atoms, 
and the number of atoms present at the critical temperature is
a few times $10^{5}$.

The magnetic field measurements also help to explain why the analysis 
of the expansion of the trapped atoms after release works so well. 
Because of the fast reversal, the atoms which make transitions to 
the $m=0$ state are indeed released extremely rapidly.
A careful analysis of the expansion may require taking into account 
the behavior of the weak field seeking atoms observed in Fig. 
\ref{fig:m1m0}. 
Here we assume that all atoms expand freely independent of their 
internal state. 
In fact the atoms in this state are presumably trapped during the 
decay of the eddy currents, but since in a cloverleaf trap, the 
confinement rapidly decreases with increasing bias, it is probably a 
good approximation to treat the atoms as free on the scale of 1 ms. 

An analysis of the mean field expansion of the cloud, using the 
corrected number of atoms leads to a value of the scattering 
length, $a=20\pm10\:\mathrm{nm}$. This result is consistent with our
elastic rate constant measurements at 1 mK\cite{Browaeys}, as well as 
with the observations of Ref.~\cite{Pereira:01b}.
 
We have also observed the ions produced by the trapped condensate,
by negatively biasing a grid above the MCP.
An example of the ion detection rate as a function of time is shown in 
Fig.~4 of Ref.~\cite{Robert:01}.
These ions are due to Penning ionization of residual gases,
to two body collisions within the condensate, or possibly other, more
complicated processes.
We observe a factor of 5 more ions from the condensate
than from a thermal cloud at 1 $\mu$K, and we attribute
this increase to the larger density in the condensate.
The lifetime of the condensate, estimated by observing the
ion rate is on the order of a few seconds.
This is true both with and without an RF-knife
to evacuate hot atoms\cite{Mewes:96,Burt},
although the lifetime is slightly longer with
the knife present.
The density of the condensate, deduced from its vertical
size measurement and its known aspect ratio, is of order
$10^{13}\,\mathrm{cm}^{-3}$,
so from the lifetime we can place an upper limit of
$10^{-13}\,\mathrm{cm}^{3}\mathrm{s}^{-1}$ on the
relaxation induced Penning ionization rate constant, 
as well as an upper limit of 
$10^{-26}\,\mathrm{cm}^{-6}\mathrm{s}^{-1}$
on any three-body loss process.

The achievement of BEC in He$^*$ together with a MCP detector,
offers many new possibilities for the investigation of BECs. 
Ion detection allows continuous
``non-destructive'' monitoring of the trapped condensate.
We hope to be able to study the formation kinetics of the condensate 
using the ion signal.
Our ability
to count individual He$^*$ atoms falling out of the trap 
should allow us to perform  accurate comparisons of
correlation functions\cite{Burt} for a thermal beam of ultracold
atoms\cite{Yasuda} and for an atom laser, realizing the quantum
atom optics counterpart of one of the fundamental experiments of
quantum optics.

\section*{Acknowledgments}
This work
was supported by the European Union under grants IST-1999-11055, and
HPRN-CT-2000-00125, and by the DGA grant 99.34.050.

\end{document}